\documentclass[preprint,1p,times]{elsarticle}
\usepackage{lineno} 
\usepackage{graphicx} 
\usepackage{amsmath}
\usepackage{amssymb}
\usepackage{graphicx}
\usepackage{lineno}
\usepackage{hyperref}
\usepackage{textcomp}
\usepackage{natbib}
\journal{Nuclear Instruments and Methods in Physics Research Section A}

\begin{document}

\begin{frontmatter}

\title{Simulation Study for Particle Identification with the dRICH of the ePIC Experiment at the EIC}

\author[1,2]{Tiziano Boasso}
\author[1]{Chatterjee Chandradoy}
\author[1]{Dalla Torre Silvia}
\author[1,2]{Martin Anna}
\author[1]{Tessarotto Fulvio}
\author[1]{Agarwala Jinky}
\author[3]{Contalbrigo Marco}
\author[3]{Polizzi Lorenzo}
\author[4,4bis]{Occhiuto Luisa}
\author[5,5bis]{Del Caro Annalisa}
\author[6]{Nagorna Tetiana}
\author[6]{Osipenko Mikhail}
\author[6]{Vallarino Simone}
\author[7]{Farokhi Fateme}
\author[8]{Kiselev Alexander}
\author[9]{Bhadauria Rohit Singh}
\author[9]{Ghosh Tapasi}
\author[10]{Nunez Cynthia}
\author[10]{Pecar Connor}
\author[11]{George Nebin}
\author[11]{Rajan Adithyan}
\author[11]{Samuel Deepak}
\author[12]{Jangid Rohit}
\author[12]{Kumar Ramandeep}
\author[12]{Laishram Girdish}
\author[12]{Tanvi Tanya}
\author{Thakur Meenu$\mathrm{^{n}}$ on behalf of the ePIC Dual RICH subsystem Collaboration}

\address[1]{INFN Sezione di Trieste, Trieste, Italy}
\address[2]{University of Trieste, Trieste, Italy}
\address[3]{INFN Sezione di Ferrara, Ferrara, Italy}
\address[4]{INFN Laboratori Nazionali di Frascati, Gruppo Collegato di Cosenza, Cosenza, Italy}
\address[4bis]{University of Cosenza, Cosenza, Italy}
\address[5]{University of Salerno, Salerno, Italy}
\address[5bis]{INFN Sezione di Napoli, Gruppo Collegato di Salerno, Salerno, Italy}
\address[6]{INFN Sezione di Genova, Genova, Italy}
\address[7]{INFN Laboratori Nazionali del Sud, Catania, Italy}
\address[8]{Brookheav National Laboratory, Upton, USA }
\address[9]{Ramaiah University, Bangalore, India}
\address[10]{Duke University, Durham, USA}
\address[11]{Central University of Karnataka, Gulbarga, India}
\address[12]{Central University of Haryana, Mahendragarh, India}

\begin{abstract}
The dual-radiator Imaging Cherenkov detector (dRICH) is a key component of the forward particle identification system for the ePIC experiment at the Electron-Ion Collider (EIC). This study evaluates the dRICH performance using Geant4 simulations in the context of the global ePIC simulation stack, focusing on the optimization of the aerogel radiator and the impact of sensor noise. We compare two aerogel configurations: the initial design (n=1.019) and the current default (n=1.026). The latter, characterized by improved optical properties and a higher refractive index, demonstrates enhanced $\pi-K$ separation at high momenta, effectively extending the operational overlap with the $\mathrm{C_2F_6}$ gas radiator. Additionally, the study investigates the impact of Silicon Photomultiplier (SiPM) dark noise, showing that a 300 kHz noise rate per channel leads to a moderate reduction (approximately 1.5 GeV/c) in the $3\sigma$ separation threshold. These results validate the current dRICH design and quantify the purity levels achievable for both radiators under expected experimental conditions.

\end{abstract}
\begin{keyword}
ePIC \sep RICH \sep radiator gas \sep aerogel \sep simulation
\end{keyword}
\end{frontmatter}

\section{Introduction}
The ePIC experiment at the Electron–Ion Collider (EIC) is an approved project under realization at Brookhaven National Laboratory (BNL), USA. It aims to address key questions on how nucleon properties such as mass and spin emerge from partons and their interactions, and how partons are distributed in momentum and position space. The EIC will also study the interactions of color-charged quarks, gluons, and jets with nuclear matter. Central questions include how confined hadronic states emerge from quarks and gluons, how quark–gluon interactions generate nuclear binding, how the nuclear environment modifies parton dynamics and correlations, and how gluon densities in nuclei evolve toward saturation at high energies, potentially giving rise to a universal gluonic phase of matter in nuclei and nucleons.
To explore such elaborate physics, the requirements on the accelerator and the detector are state-of-the-art. High polarization ($\sim$ 70\%) for electron and light nuclei beams, availability of heavy ions up to Uranium, high luminosity (peak value around 10$^{33}$ to 10$^{34}$ cm$^2$s$^{-1}$), high centre of mass energy (20-140 GeV) makes the EIC an ultimate machine to study the fundamental aspect of QCD and hadronic physics. The detector has the general purpose to deliver the whole EIC physics program; almost hermetic coverage and compact setup are key challenges for such multipurpose ambitious detector design. The requirements for EIC have been described in detail in the common effort by the physics community interested in the EIC physics, known as the Yellow report \cite{Khalek:2021aa}. In the Yellow report the role of the particle identification (PID) detectors has been considered fundamental in delivering the physics of EIC. Several technologies have been identified to deliver the PID requirements altogether. A dual radiator Ring Imaging CHerenkov (dRICH) counter in the forward direction will provide at three sigma level (i.e. 3$\sigma$, three times the Cherenkov angle resolution), kaon-pion separation in a pseudorapidity range between 1.5 to 3.5. Two radiators will provide a continuos PID from few GeV/c up to 50 GeV/c. It will support the identification of the scattered electrons of Deep Inelastic Scattering (DIS) events by pion rejection, therefore complementing the calorimeter systems, particularly at low momenta, up to 15 GeV/c. The aerogel radiator will provide particle identification in the lower momentum end and the $\mathrm{C_2F_6}$ gaseous radiator  in the higher momentum range up to 50 GeV/c, with substantial overlap between the two ranges. The RICH design is based on six identical arrangements of spherical mirrors and photosensor sectors where SiPMS sensors, insensitive the the magnetic field of the ePIC spectrometer, are arranged to form a pseudospherical surface. Each mirror focuses reflected photons onto the corresponding photosensor area which approximates the mirror focal area. A sophisticated simulation serves as a fundamental block of the detector. For the ePIC dRICH an extensive simulation campaign is ongoing. In this study, we discuss the results of several simulation studies performed with the official dRICH simulation stack and using single particle Monte-Carlo events.
\section{Simulation software scheme}
\begin{figure}[!thb]
    \centering
    \includegraphics[width=0.70\linewidth]{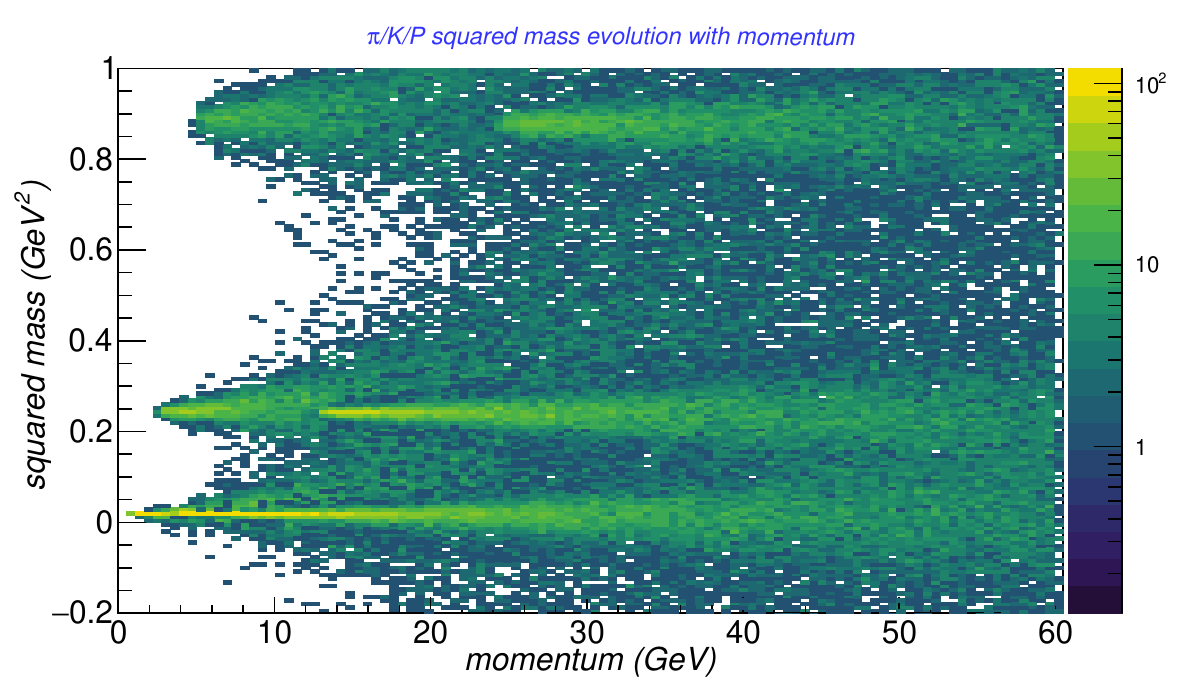}
    \caption{Reconstructed $m^2$ as a function of momentum for single particle events in the dRICH simulation environment. The plot was produce using the n=1.019 aerogel. }
    \label{fig:m2p}
    \vspace{-0.2cm}
\end{figure}

The ePIC software scheme is complex and multi-layered. To simulate the hits, the geometry of the detector is described in the DD4hep framework \cite{dd4hep}, and the simulated data is generated using Geant4 \cite{AGOSTINELLI2003250}; for our studies related to generation of Cherenkov photons we use the plugin \cite{npsim} of DD4hep. The simulated hits are then reconstructed using an indirect ray tracing method \cite{irt} to obtain the Cherenkov angle. This standalone code based on C++ is a part of the ePIC reconstruction framework EICRecon \cite{eicrecon}. EICRecon is a framework based on JANA2 \cite{jana}. The data model of the simulation studies strictly follows the data model of the ePIC experiment \cite{edm4eic}, which  is based on PODIO \cite{podio} and EDM4hep\cite{EDM4hep}.
The final output for the characterization studies of the sub-detector systems is fully customized by the users. For example, in figure \ref{fig:m2p} it can be seen how the reconstructed squared mass of charged particles varies as a function of momentum. The combined information provided by the two radiators can be appreciated. The mass has been computed using the dRICH Cherenkov angle information. The simulation framework has also been extensively utilized for some the critical parameter optimizations and impact studies, as those reported here. Several other important studies are being performed and have been reported elsewhere \cite{Chatterjee:2024aa}. 
\section{Improving Aerogel Performance}
\begin{figure}[h]
    \centering
    \includegraphics[width=0.8\linewidth]{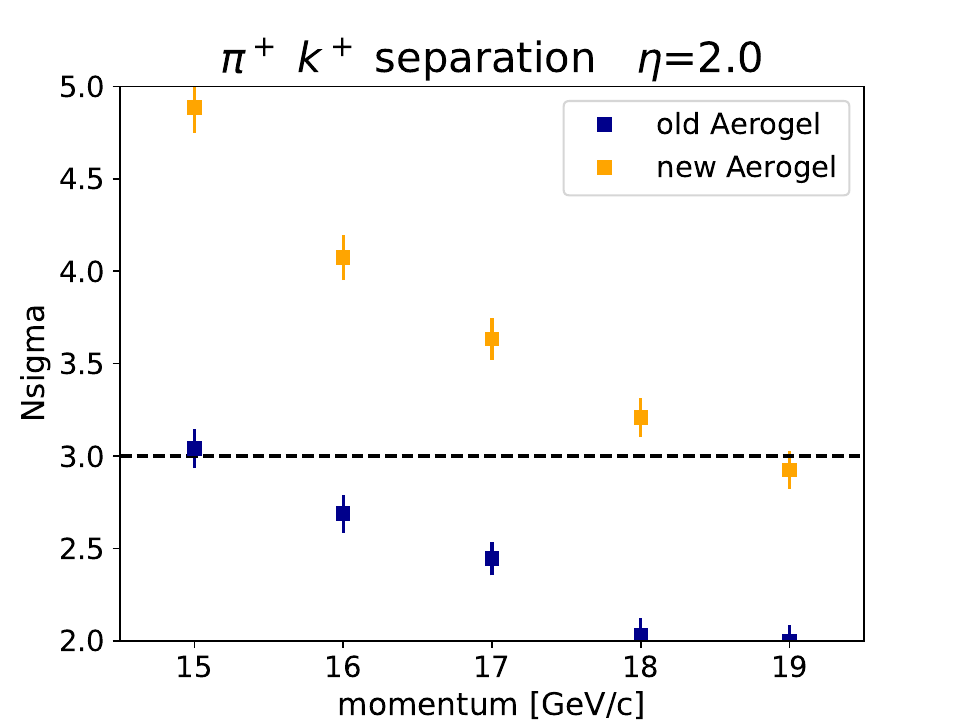}
    \caption{$N\sigma\ \pi-K$ curve for the two aerogel; $n=1.019$ is labels as "old" and the current default as "new". }
    \label{fig:oldnew}
\end{figure}

The aerogel radiator of the dRICH provides particle identification (PID) at low momentum, specifically where the charged particles are below the Cherenkov threshold of the gas radiator. The threshold for kaon identification in the $\mathrm{C_2F_6}$ radiator, as measured in laboratory characterization and in beam tests with prototypes, is approximately 12 GeV/c. Therefore, to ensure substantial overlap with the gas radiator, the aerogel must provide $3\sigma$ separation well above 12 GeV/c. In this study, the parameters of two aerogel samples were used for a comparative performance analysis. The first sample, originally planned for the ePIC experiment, has an average refractive index of 1.019\cite{VALLARINO2024168834}; the second, with an index of 1.026, is the new current default for the detector. The radiator is planned to be 4 cm thick. An aerogel radiator with a higher refractive index increases the number of detected photons, thereby enhancing pattern recognition; however, saturation occurs at a lower momentum.
The new aerogel features improved optical properties, with an optimized Rayleigh scattering length that improves the resolution of the reconstructed single-photoelectron Cherenkov angle. Overall, the aerogel allows for better particle discrimination at higher momenta, extending the overlapping range with the gas. In figure \ref{fig:oldnew} the separation $\pi-K$ in number of standard deviation $N\sigma$ is plotted as a function of momentum at a fixed pseudorapidity ($\eta=2.0$). 
The number of detected photons is also heavily influenced by the pseudorapidity due to geometric acceptance: at $\eta=2.0$, the ring is typically fully contained within a single sensor sector. At higher values of $\eta$, performance is modulated by the "shadowing" effect of the beam pipe and the splitting of rings across multiple sensor sectors as they are reflected by nearby mirrors.
Increasing the number of detected photons improves pattern recognition and leads to better ring resolution. For this reason, we investigated how the $N\sigma$ curve changed when increasing the aerogel thickness up to 6 cm. No self-absorption effects were observed, and performance clearly improved; however, due to construction complexities, there are currently no plans to increase the radiator thickness.
The simulation is in agreement with the studies performed in laboratory and a series of beam test studies performed at CERN \cite{marco_RICH25}. 

\section{Noise Impact}
\begin{figure}[h]
    \centering
\includegraphics[width=0.8\linewidth]{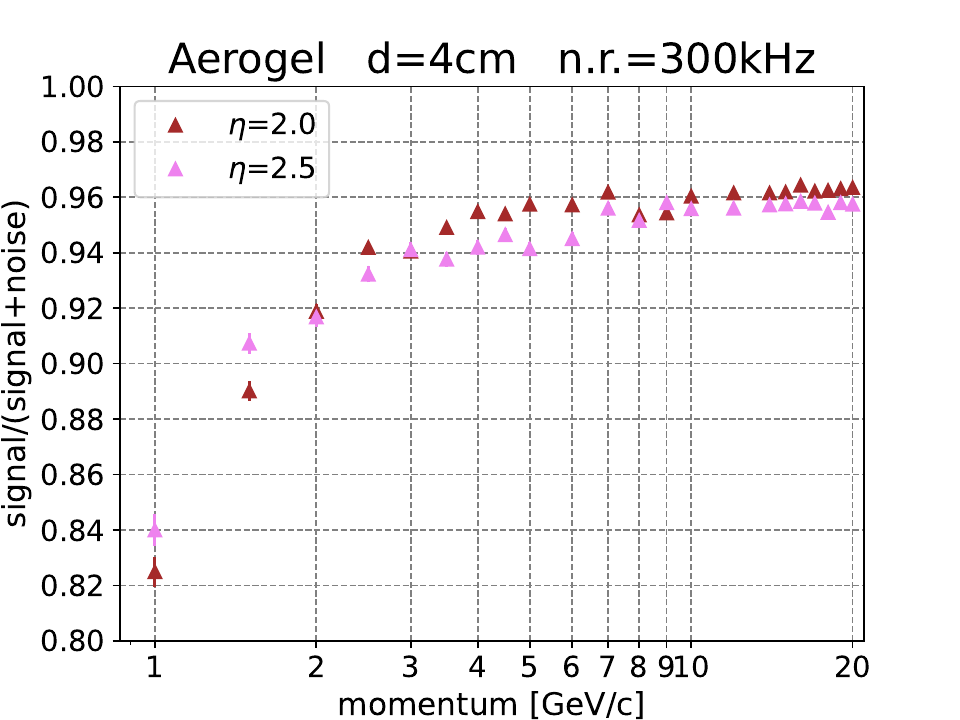}
    \caption{Signal purity as a function of momentum for the aerogel ring. It increase up to saturation resembling the photon yield. }
    \label{fig:purity}
\end{figure}

\begin{figure}[h]
    \centering
\includegraphics[width=0.8\linewidth]{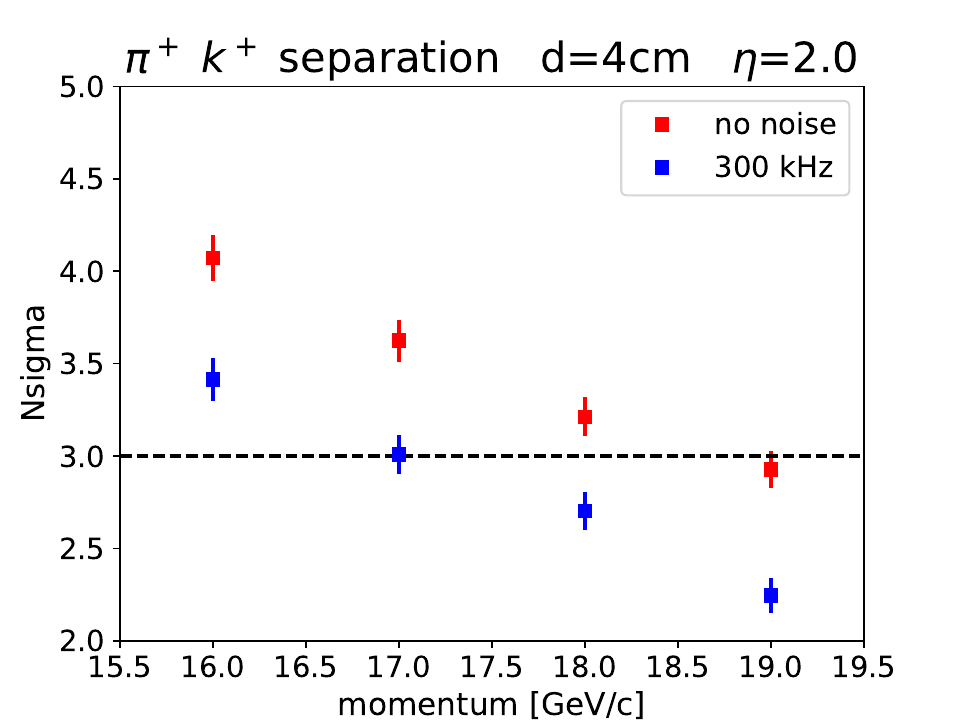}
    \caption{$N\sigma$ curve with and without simulating the noise.}
    \label{fig:noise}
\end{figure}

SiPMs are chosen as baseline photosensor for ePIC, thanks to the several advantages of this technology: magnetic field insensitivity, high photo-detection efficiency, high granularity, cost, high timing performance and the requirement of low bias voltage. At the same time, there is  a concerning challenge. The intrinsic Dark Count Rate (DCR) is high and largely increased with high radiation doses. The ePIC dRICH collaboration has a mitigation strategy to this problem \cite{PREGHENELLA2023167661}, however, after 5 years of operation in ePIC at the highest EIC luminosity, a maximum noise rate of 300 kHz/channel is expected. In each dRICH sector, this correspond to 50 DCR hits over 50 K pixels within a 3 ns event gating. A particular concern is the contamination of DCR hits to the aerogel ring, due to its large Cherenkov angle. To simulate white noise, counts are added uniformly across the sensor surface before reconstruction assuming a time window of 1 ns. Purity is defined as $Signal / (Signal + Noise)$. To extract the noise contribution in the signal region, a combined fit utilizing a Gaussian for the signal and a first-order polynomial for the noise was applied. The noise was calculated as the integral of the polynomial curve within a $3\sigma$ window of the Gaussian peak, while the signal was determined by subtracting this estimated noise from the total counts. $\mathrm{C_2F_6}$ ring is smaller because its refractive index \cite{ABJEAN1995417} is lower than that of the aerogel, but the photon count is higher due to the length of the gas volume. This results in a purity of approximately $99\%$ for the $\mathrm{C_2F_6}$ ring, while for the aerogel, it reaches only about $96\%$ as shown in figure \ref{fig:purity}, degradating the aerogel ring resolution. As shown in figure \ref{fig:noise}, the PID performance of the new aerogel with added noise shows that the $3\sigma$ $\pi-K$ separation limit decreases by approximately 1.5 GeV/c compared to cases without noise injection.
\section{Conclusions}
The comprehensive simulation of the dRICH detector demonstrates that the current design meets the PID requirements of EIC and that a substantial overlap between the two radiators is achievable even considering the dark noise of the SiPMs.
\section*{Acknowledgments}
    This work was carried out within the ePIC-dRICH Collaboration. It was funded by INFN, Italy, and supported by US DOE. The authors thank the ePIC Collaboration and EIC project, the technical and management staff at INFN, BNL and JLab laboratories.

\footnotesize
\bibliography{bib}
\end{document}